\documentclass[conference]{IEEEtran-v17}

\pdfoutput=1

\newlength{\figwidth}
\usepackage[nosort]{cite}
\usepackage{url}
\usepackage[intlimits]{amsmath}
\usepackage{bbm}
\usepackage{graphicx}
\usepackage{paralist}
\usepackage{fancyref}
\usepackage[stretch=16,shrink=16,step=4]{microtype}
\usepackage{bm}
\usepackage{upgreek}
\usepackage{amssymb}
\usepackage{amsfonts}
\usepackage{mathrsfs}
\usepackage{xspace}
\usepackage{bbold}

\makeatletter
\def\amsbb{\use@mathgroup \M@U \symAMSb}
\makeatother

\newcommand{\lefto}{\mathopen{}\left}
\newcommand{\safemath}[2]{\newcommand{#1}{\ensuremath{#2}\xspace}}
\safemath{\opE}{\amsbb{E}}
\newcommand{\Ex}[2]{\ensuremath{\amsbb{E}_{#1}\lefto[#2\right]}}
\safemath{\prob}{\amsbb{P}}
\safemath{\bigO}{\mathcal{O}}
\safemath{\littleo}{\mathit{o}}
\newcommand{\diag}[1]{\mathrm{diag}\left( #1 \right)}

\newcommand{\herm}[1]{\ensuremath{#1^{\mathsf{H}}}} 	% hermitian transpose
\newcommand{\tr}[1]{\mathrm{tr} \lefto( \ensuremath{#1} \right )}

\newcommand{\deter}[1]{\mathrm{det}\lefto( \ensuremath{#1} \right)}
\newcommand{\cgauss}[2]{\mathcal{CN} \lefto( \ensuremath{#1, #2}  \right) }

\newcommand{\gamfunc}[2]{\Gamma_{#1} \lefto( \ensuremath{#2}  \right) }

\newcommand{\subtext}[1]{\text{\fontfamily{cmr}\fontshape{n}\fontseries{m}\selectfont{}#1}}
\newcommand{\sub}[1]{\ensuremath{_{\subtext{#1}}}} % for text mode subscripts
\safemath{\txant}{M\sub{t}}
\safemath{\rxant}{M\sub{r}}

\safemath{\cohtime}{\ensuremath{T}}
\safemath{\codeblocks}{\ensuremath{L}}

\newtheorem{thm}{Theorem}

\newtheorem{rem}{Remark}
\newtheorem{prop}[thm]{Proposition}
\safemath{\matA}{\mathsf{A}}
\safemath{\matB}{\mathsf{B}}
\safemath{\matC}{\mathsf{C}}
\safemath{\matD}{\mathsf{D}}
\safemath{\matE}{\mathsf{E}}
\safemath{\matF}{\mathsf{F}}
\safemath{\matG}{\mathsf{G}}
\safemath{\matH}{\mathsf{H}}
\safemath{\matI}{\mathsf{I}}
\safemath{\matJ}{\mathsf{J}}
\safemath{\matK}{\mathsf{K}}
\safemath{\matL}{\mathsf{L}}
\safemath{\matM}{\mathsf{M}}
\safemath{\matN}{\mathsf{N}}
\safemath{\matO}{\mathsf{O}}
\safemath{\matP}{\mathsf{P}}
\safemath{\matQ}{\mathsf{Q}}
\safemath{\matR}{\mathsf{R}}
\safemath{\matS}{\mathsf{S}}
\safemath{\matT}{\mathsf{T}}
\safemath{\matU}{\mathsf{U}}
\safemath{\matV}{\mathsf{V}}
\safemath{\matW}{\mathsf{W}}
\safemath{\matX}{\mathsf{X}}
\safemath{\matY}{\mathsf{Y}}
\safemath{\matZ}{\mathsf{Z}}
\safemath{\matSigma}{\mathsf{\Sigma}}
\safemath{\matLambda}{\mathsf{\Lambda}}
\safemath{\matDelta}{\mathsf{\Delta}}

\safemath{\randveca}{\bm{A}}
\safemath{\randvecb}{\bm{B}}
\safemath{\randvecc}{\bm{C}}
\safemath{\randvecd}{\bm{D}}
\safemath{\randvece}{\bm{E}}
\safemath{\randvecf}{\bm{F}}
\safemath{\randvecg}{\bm{G}}
\safemath{\randvech}{\bm{H}}
\safemath{\randveci}{\bm{I}}
\safemath{\randvecj}{\bm{J}}
\safemath{\randveck}{\bm{K}}
\safemath{\randvecl}{\bm{L}}
\safemath{\randvecm}{\bm{M}}
\safemath{\randvecn}{\bm{N}}
\safemath{\randveco}{\bm{O}}
\safemath{\randvecp}{\bm{P}}
\safemath{\randvecq}{\bm{Q}}
\safemath{\randvecr}{\bm{R}}
\safemath{\randvecs}{\bm{S}}
\safemath{\randvect}{\bm{T}}
\safemath{\randvecu}{\bm{U}}
\safemath{\randvecv}{\bm{V}}
\safemath{\randvecw}{\bm{W}}
\safemath{\randvecx}{\bm{X}}
\safemath{\randvecy}{\bm{Y}}
\safemath{\randvecz}{\bm{Z}}

\safemath{\randmatA}{\amsbb{A}}
\safemath{\randmatB}{\amsbb{B}}
\safemath{\randmatC}{\amsbb{C}}
\safemath{\randmatD}{\amsbb{D}}
\safemath{\randmatE}{\amsbb{E}}
\safemath{\randmatF}{\amsbb{F}}
\safemath{\randmatG}{\amsbb{G}}
\safemath{\randmatH}{\amsbb{H}}
\safemath{\randmatI}{\amsbb{I}}
\safemath{\randmatJ}{\amsbb{J}}
\safemath{\randmatK}{\amsbb{K}}
\safemath{\randmatL}{\amsbb{L}}
\safemath{\randmatM}{\amsbb{M}}
\safemath{\randmatN}{\amsbb{N}}
\safemath{\randmatO}{\amsbb{O}}
\safemath{\randmatP}{\amsbb{P}}
\safemath{\randmatQ}{\amsbb{Q}}
\safemath{\randmatR}{\amsbb{R}}
\safemath{\randmatS}{\amsbb{S}}
\safemath{\randmatT}{\amsbb{T}}
\safemath{\randmatU}{\amsbb{U}}
\safemath{\randmatV}{\amsbb{V}}
\safemath{\randmatW}{\amsbb{W}}
\safemath{\randmatX}{\amsbb{X}}
\safemath{\randmatY}{\amsbb{Y}}
\safemath{\randmatZ}{\amsbb{Z}}
\safemath{\randmatSigma}{\mathbb{\Sigma}}
\safemath{\randmatPhi}{\mathbb{\Phi}}
\safemath{\randmatDelta}{\mathbb{\Delta}}

\safemath{\pdff}{f}
\safemath{\pdfp}{p}
\safemath{\pdfq}{q}

\safemath{\cdfF}{F}
\safemath{\cdfP}{P}
\safemath{\cdfQ}{Q}

\safemath{\veca}{\bm{a}}
\safemath{\vecb}{\bm{b}}
\safemath{\vecc}{\bm{c}}
\safemath{\vecd}{\bm{d}}
\safemath{\vece}{\bm{e}}
\safemath{\vecf}{\bm{f}}
\safemath{\vecg}{\bm{g}}
\safemath{\vech}{\bm{h}}
\safemath{\veci}{\bm{i}}
\safemath{\vecj}{\bm{j}}
\safemath{\veck}{\bm{k}}
\safemath{\vecl}{\bm{l}}
\safemath{\vecm}{\bm{m}}
\safemath{\vecn}{\bm{n}}
\safemath{\veco}{\bm{o}}
\safemath{\vecp}{\bm{p}}
\safemath{\vecq}{\bm{q}}
\safemath{\vecr}{\bm{r}}
\safemath{\vecs}{\bm{s}}
\safemath{\vect}{\bm{t}}
\safemath{\vecu}{\bm{u}}
\safemath{\vecv}{\bm{v}}
\safemath{\vecw}{\bm{w}}
\safemath{\vecx}{\bm{x}}
\safemath{\vecy}{\bm{y}}
\safemath{\vecz}{\bm{z}}

\safemath{\setA}{\mathcal{A}}
\safemath{\setB}{\mathcal{B}}
\safemath{\setC}{\mathcal{C}}
\safemath{\setD}{\mathcal{D}}
\safemath{\setE}{\mathcal{E}}
\safemath{\setF}{\mathcal{F}}
\safemath{\setG}{\mathcal{G}}
\safemath{\setH}{\mathcal{H}}
\safemath{\setI}{\mathcal{I}}
\safemath{\setJ}{\mathcal{J}}
\safemath{\setK}{\mathcal{K}}
\safemath{\setL}{\mathcal{L}}
\safemath{\setM}{\mathcal{M}}
\safemath{\setN}{\mathcal{N}}
\safemath{\setO}{\mathcal{O}}
\safemath{\setP}{\mathcal{P}}
\safemath{\setQ}{\mathcal{Q}}
\safemath{\setR}{\mathcal{R}}
\safemath{\setS}{\mathcal{S}}
\safemath{\setT}{\mathcal{T}}
\safemath{\setU}{\mathcal{U}}
\safemath{\setV}{\mathcal{V}}
\safemath{\setW}{\mathcal{W}}
\safemath{\setX}{\mathcal{X}}
\safemath{\setY}{\mathcal{Y}}
\safemath{\setZ}{\mathcal{Z}}
\safemath{\emptySet}{\varnothing}

\safemath{\veczero}{\mathbf{0}} %vector font of 0,

\safemath{\jpg}{\mathcal{CN}}			% jointly proper Gaussian
\safemath{\complexset}{\amsbb{C}}
\safemath{\realset}{\amsbb{R}}

\usepackage{bm}
\usepackage{upgreek}
\usepackage{amssymb}
\usepackage{amsfonts}
\usepackage{mathrsfs}
\usepackage{xspace}

\safemath{\mi}{I}
\safemath{\difent}{\mathrm{h}}		%differential entropy
\safemath{\NonnegReal}{\mathbb{R}^{+}}
\safemath{\genericpdf}{f}
\safemath{\altsnr}{\tilde{\snr}}
\safemath{\unifdist}{P^{\mathrm{U}}}
\safemath{\infoden}{\imath}
\safemath{\iso}{\mathrm{iso}}
\safemath{\funcpdf}{\varphi}

\safemath{\indist}{\cdfP} %input distribution
\safemath{\outdist}{\cdfQ} %output distribution
\safemath{\inpdf}{\pdfp} %input pdf
\safemath{\outpdf}{\pdfq} %output pdf
\safemath{\testdist}{\cdfP} %the distribution of test P_{Z|W}

\safemath{\powallocvec}{\vecv}
\safemath{\powalloc}{v}
\safemath{\encoder}{f} % the encoder of the code
\safemath{\decoder}{g} % the decoder of the code
\safemath{\msg}{J} % the message
\safemath{\csir}{\mathrm{r}}
\safemath{\csit}{\mathrm{t}}
\safemath{\csi}{\mathrm{csi}}
\safemath{\csirt}{\mathrm{rt}}
\safemath{\Rcsirt}{R_{\csirt}} %rate with CSIT
\safemath{\Rcsir}{R_{\csir}}
\safemath{\Rcsit}{R_{\csit}}

\safemath{\Rnocsit}{R_{\mathrm{no}}} %rate without CSIT
\safemath{\Rnocsi}{R_{\mathrm{no}}} %rate without CSIT
\safemath{\Rgeneral}{R_{\mathrm{no,rt}}} %rate with general CSI condition
\safemath{\equalR}{R_{\mathrm{e}}}

\safemath{\Rmax}{R^*(n,\epsilon)} %rate without CSIT
\safemath{\codewords}{K}
\safemath{\Rub}{\bar{R}(n,\epsilon)}
\safemath{\Rlb}{\underline{R}(n,\epsilon)}

\safemath{\randtemp}{K}

\safemath{\Vnocsit}{V_{\error}^{\mathrm{no}}}
\safemath{\Vcsit}{V_{\error}^{\mathrm{rt}}}

\safemath{\cadist}{F_C} %capacity distribution
\safemath{\cdistno}{F_{0}}
\safemath{\cdistcsit}{F_{1}}

\safemath{\deF}{d_0} %the derivative of F_C(\xi) %\left.\frac{d\cadist(\argpn)}{d\argpn}\right|_{\argpn=C_\error}

\safemath{\bl}{n} %block length
\safemath{\blt}{\tilde{\bl}}
\safemath{\error}{\epsilon} %prob of error
\safemath{\NumCode}{M}
\safemath{\RXant}{r}
\safemath{\txantop}{t^{\ast}}
\safemath{\minant}{m}
\safemath{\minantop}{\minant^\ast}
\safemath{\snr}{\rho}
\safemath{\const}{k}
\safemath{\inset}{\setF} %the input power set
\safemath{\spanm}{\mathrm{span}}
\safemath{\ubb}{b_{0}}%upper bound of b_1,...,b_{r} of the matrix I+ H^H U H.

\safemath{\covmat}{\matU}
\safemath{\covmatY}{\matV}
\safemath{\randcmY}{\randmatV}

\safemath{\insetMIMO}{\setF_{\txant,\bl}}
\safemath{\insetcov}{\setU_{\txant}}
\safemath{\condcov}{\matQ\in\insetcov}

\safemath{\BoundFU}{k_{\delta}}

\safemath{\argpn}{\xi} % the argument in the probability p(R)

\safemath{\angletest}{Z} %the angle test used in achievability part

\safemath{\Cnocsit}{C_{\epsilon}^{(0)}}
\safemath{\Ccsit}{C_{\epsilon}^{(1)}}

\safemath{\errorach}{P_\mathrm{e}} %the error in intuition part

\newcommand{\given}{\,\vert\,}				% conditioning
	
\safemath{\define}{\triangleq}			% definition
\safemath{\altbl}{\tilde{\bl}}
\safemath{\constL}{k_{\mathrm{L}}}
\safemath{\constU}{k_{\mathrm{U}}}
\safemath{\funcL}{\tilde{q}}
\safemath{\funcU}{q}
\safemath{\ConstThm}{k_0}
\safemath{\randrevec}{\randvecy}
\safemath{\revec}{\vecy}
\safemath{\trcwd}{\vecx}
\safemath{\randtrcwd}{\randvecx}
\safemath{\randnoisevec}{\randvecw}
\safemath{\transmitcwd}{\vecx_1} %the transmitted codeword in achievability
\safemath{\pickcwdnoch}{\vecx_0} %the chosen codeword x=[P,P,...,P]
\safemath{\pickcwd}{\vecx_0} %the chosen codeword x=[P,P,...,P]

\safemath{\inseqrand}{\randvecx}
\safemath{\inseq}{\vecx}

\safemath{\outseq}{\matY}
\safemath{\outseqrand}{\randmatY}

\safemath{\altT}{\widetilde{T}}
\safemath{\altU}{\widetilde{U}}
\safemath{\altmean}{\tilde{\mu}}
\safemath{\altvar}{\tilde{\sigma}}
\safemath{\altf}{\funcL}
\safemath{\altg}{\tilde{g}}
\safemath{\altgamma}{\tilde{\gamma}}
\safemath{\altdelta}{\tilde{\delta}}
\safemath{\altk}{\tilde{k}}
\safemath{\altconst}{\altk}
\safemath{\altbmLambda}{\widetilde{\bm{\Lambda}}}
\safemath{\constant}{\tilde{k}}

\makeatletter
\def\@IEEEinterspaceratioM{0.265}
\def\@IEEEinterspaceMINratioM{0.1651}
\def\@IEEEinterspaceMAXratioM{0.38}

\def\@IEEEinterspaceratioB{0.31}
\def\@IEEEinterspaceMINratioB{0.19}
\def\@IEEEinterspaceMAXratioB{0.38}
\@IEEEtunefonts
\makeatother
\begin{document}

\IEEEoverridecommandlockouts
% DRAFT
% to include revision information into the resulting PDF
%%%%\svnInfo $Id: dbs_it07.tex 2697 2009-08-04 10:43:20Z gdurisi $
%% add a PDFinfo field to store metadata in the output PDF

%\IEEEpubid{978-1-4799-5863-4/14/\$31.00 \copyright 2014 IEEE}

% paper title
\title{Diversity versus Multiplexing at Finite Blocklength}

\author{\IEEEauthorblockN{Johan \"Ostman$^1$, Wei Yang$^1$, Giuseppe Durisi$^1$,  Tobias Koch$^2$}
\\
\IEEEauthorblockA{
$^1$Chalmers University of Technology, 41296 Gothenburg, Sweden\\
$^2$Universidad Carlos III de Madrid, 28911 Legan\'{e}s, Spain\\
}}

%
%
% make the title area			
\maketitle

%%%%%%%%%%%%%%%%
\begin{abstract}
  A finite blocklenth analysis of the diversity-multiplexing tradeoff is presented, based on nonasymptotic bounds on the maximum channel coding rate of  multiple-antenna block-memoryless Rayleigh-fading channels.
  The bounds in this paper allow one to numerically assess for  which packet size, number of antennas, and degree of channel selectivity, diversity-exploiting schemes are close to optimal, and when instead the available spatial degrees of freedom should be used to provide spatial multiplexing.
This finite blocklength view on the diversity-multiplexing tradeoff provides insights on the design of delay-sensitive ultra-reliable communication
links.
  \renewcommand{\thefootnote}{}
\footnote{This research was  supported by the Swedish Research Council under grant 2012-4571,  by a Marie Curie FP7 Integration Grant within the 7th European Union Framework Programme under Grant 333680, and by the Spanish Government (CSD2008-00010, TEC2009-14504-C02-01, and TEC2013-41718-R).}
\end{abstract}
\setcounter{footnote}{0}

%%%%%%%%%%%%%%%%%%%%%%%%%%%%%%%%%%
\section{Introduction}
Multi-antenna technology is by now a fundamental part of all modern wireless communication standards, due to its ability to provide impressive gains in both spectral efficiency and reliability.  
Multiple antennas yield additional spatial degrees of freedom that can be used to lower the error probability for a given rate, through the exploitation of \emph{spatial diversity}, or increase the rate for a given error probability, through the exploitation of \emph{spatial multiplexing}.
These two effects cannot be harvested concurrently and there exists a fundamental tradeoff between diversity and multiplexing.
This tradeoff admits a particularly simple characterization in the regime of high signal-to-noise ratio (SNR) and infinitely long data packets (codewords)~\cite{zheng03-05a}.

Current cellular systems operate typically at maximum multiplexing~\cite{lozano10-09a}. 
Indeed, diversity-exploiting techniques such as space-time codes turn out to be detrimental for low-mobility users, for which the fading coefficients can be learnt easily at the transmitter and outage events can be avoided altogether by rate adaptation.
On the other hand, for high-mobility users, diversity-exploiting techniques are not advantageous because of the abundant time and frequency selectivity that is available.
These considerations hold under the assumptions of long packet lengths ($1000$ channel uses or more) and moderately low packet error probability (around $10^{-2}$).

As we move towards next generation wireless communication systems (5G), these assumptions may cease to be valid.
Emerging applications in 5G (such as metering, traffic safety, and telecontrol of industrial plants) may require the exchange of short packets, sometimes under stringent latency and reliability constraints~\cite{metis-project-deliverable-d1.113-04a,boccardi14-02a}.
The question addressed in this paper is how multiple antennas should be used in this scenario. Is  diversity more beneficial than multiplexing in the regime of short packet length (say $100$ channel uses, roughly equal to a LTE resource block) and high reliability (packet error rate equal to $10^{-5}$ or lower)? What is the cost of learning the fading coefficients, which is required to exploit spatial degrees of freedom, when the packet size is short?  Does this cost overcome the benefits of multiple antennas?\footnote{In the infinite blocklength regime, this question was addressed in \cite{lapidoth02-05a,hassibi03-04a}.}
In this paper, we shed light on these questions leveraging on recent progresses in finite blocklength information theory~\cite{polyanskiy10-05a,yang14-07c,yang12-09a}.
%\IEEEpubidadjcol 

\paragraph*{Notation} % (fold)
\label{par:notation}
Upper case letters such as $X$ denote scalar random variables and their realizations are written in lower case, e.g., $x$. We use boldface upper case letters to denote random vectors, e.g., $\randvecx$, and boldface lower case letters for their realizations, e.g., $\vecx$. Upper case letters of two special fonts are used to denote deterministic matrices (e.g., $\matY$) and random matrices (e.g., $\randmatY$). The superscript $^H$ stands for Hermitian transposition and we use $\tr{\cdot}$ and $\deter{\cdot}$ to denote the trace and the determinant of a given matrix, respectively. 
The identity matrix of size $a\times a$ is written as $\matI_a$. 
The distribution of a circularly symmetric complex Gaussian random variable with variance $\sigma^2$ is denoted by 
$\cgauss{0}{\sigma^2}$.
Finally, $\ln(\cdot)$ indicates the natural logarithm, $[a]^+ $ stands for $\max\lbrace a,0\rbrace$, and $\gamfunc{\alpha}{\beta}$ denotes the complex multivariate Gamma function.

\section{Channel Model and Performance Metrics} % (fold)
\label{sec:system_model}
We consider a Rayleigh block-fading channel with~$\txant$ transmit antennas and~$\rxant$ receive antennas that stays constant for $\cohtime$ channel uses.
For a frequency-flat narrowband channel, $\cohtime$ is the number of channel uses in time over which the channel stays constant (coherence time); for a frequency-selective channel and under the assumption that orthogonal frequency-division multiplexing (OFDM) is used, $\cohtime$ is the number of subcarriers over which the channel stays constant (coherence bandwidth). More generally,~$\cohtime$ can be interpreted as the number of ``time-frequency slots'' over which the channel does not change.
Within the $l$th coherence interval, the channel input-output relation can be written as
\begin{IEEEeqnarray}{rCl}
    \label{eq:channel_model}
    \randmatY_l = \sqrt{{\rho}/{\txant}} \matX_l \randmatH_l + \randmatW_l.
\end{IEEEeqnarray}
Here, $\matX_l \in \complexset^{\cohtime\times \txant}$ and $\randmatY_l \in \complexset^{\cohtime \times \rxant}$ are the transmitted and received matrices, respectively; the entries of the complex fading matrix $\randmatH_l \in \complexset^{\txant\times \rxant}$ are independent and identically distributed (i.i.d.) $\cgauss{0}{1}$; $\randmatW_l \in \complexset^{\cohtime \times \rxant}$ denotes the additive noise at the receiver and has i.i.d.\ $\cgauss{0}{1}$ entries.
We assume $ \left\lbrace \randmatH_l \right\rbrace$ and $ \left\lbrace \randmatW_l \right\rbrace$ to take on independent realizations over successive coherence intervals. We further assume that $\randmatH_l$ and $\randmatW_l$ are independent and that their joint law does not depend on $\matX_l$. 

Throughout the paper, we shall focus on the so called \emph{noncoherent setting} where both the transmitter and the receiver know the distribution of $\randmatH_l$ but not its realization. 
In other words, \emph{a priori} channel state information (CSI) is not available at transmitter and at the receiver.

The assumption of no CSI at the transmitter is reasonable in a high-mobility scenario, where the fast channel variations make channel tracking at the transmitter unfeasible. 
It is also appropriate for transmission over control channels or for time-critical applications.
Indeed, in both situations it is desirable to avoid the creation of a feedback link, required to provide CSI at the transmitter.
The assumption of no \emph{a priori} CSI at the receiver allows one to characterize the information-theoretic cost of learning the channel at the receiver (for example by pilot transmission followed by channel estimation); see also \cite[Sec.~I]{yang12-09a}. This cost may be relevant when the packet size is limited and the channel is rapidly varying.

We consider coding schemes with blocklength $n=LT$.
Specifically, each codeword $\matX^L\define[\matX_1,\dots,\matX_L]$ consists of~$L$ subcodewords $\{\matX_l\}_{l=1}^{\codeblocks}$, each one undergoing a different fading realization. 
We shall refer to $L$ as the number of time-frequency diversity branches. It is a measure of the degree of time-frequency selectivity of the propagation channel.
Furthermore, each codeword is subject to the power constraint
\begin{IEEEeqnarray}{rCl}
    \label{eq:power_constraint}
    \frac{1}{\codeblocks} \sum_{l=1}^\codeblocks \tr{\matX_l\herm{\matX_l}} = \cohtime\txant.
\end{IEEEeqnarray}
Since the variance of the entries  of $\randmatH_l$ and of~$\randmatW_l$ in (\ref{eq:channel_model}) is normalized to one, $\rho$ in~\eqref{eq:channel_model} can be thought of as the SNR at each receive antenna.

We shall constrain the subcodewords $\{\matX_l\}_{l=1}^{\codeblocks}$ to be scaled unitary matrices, i.e.,
\begin{IEEEeqnarray}{rCL}
  %\IEEEeqnarraymulticol{3}{l}{...}
  % a & = & b +c
  \herm{\matX_l}\matX_l=c \cdot \matI_{M}, \quad l=1,\dots,\codeblocks
  \label{eq:code-constraint}
\end{IEEEeqnarray}
where $c$ denotes a positive constant.
Assumption~\eqref{eq:code-constraint} forces orthogonality in space, and the transmit power to be allocated uniformly across antennas and coherence intervals.
Such a power-allocation strategy is reasonable if
\begin{inparaenum}[i)]
\item the fading process is isotropic in space and i.i.d. across coherence intervals and
\item the transmitter is not aware of the channel realizations.
\end{inparaenum}
As we shall discuss in the next section, the assumption \eqref{eq:code-constraint} is, for example, satisfied by unitary space-time modulation, which achieves the ergodic capacity at high SNR when $\cohtime\geq \txant+\rxant$~\cite{zheng02-02a,yang13-02a}.

The objective of the paper is to characterize the maximal channel coding rate $R^{*}(n,\epsilon)$ among all codes that have blocklength~$n$, that satisfy both~\eqref{eq:power_constraint} and~\eqref{eq:code-constraint}, and that are decodable with average error probability not exceeding $\epsilon$.
A formal definition of $R^{*}(n,\epsilon)$ follows along the same line as~\cite[Def.~1]{yang14-07c}.

\section{Outage and Ergodic Capacity} % (fold)
\label{sec:outage_and_ergodic_capacity}
Most of the results available in the literature can be interpreted as asymptotic characterizations of $R^{*}(n,\epsilon)$ for $n\to \infty$.
For the case when $T$ and, hence $n$, grows to infinity for a fixed~$L$, the maximum coding rate  $R^{*}(n,\epsilon)$ converges to the outage capacity~$C_{\mathrm{out},\epsilon}$~\cite[p.~2631]{biglieri98-10a},~\cite{ozarow94-05a}
  \begin{IEEEeqnarray}{rCL}\label{eq:outage_capacity}
    %\IEEEeqnarraymulticol{3}{l}{...}
    % a & = & b +c
    C_{\mathrm{out},\epsilon}=\sup\lefto\{ R : P\sub{out}(R)\leq\epsilon\right\} \IEEEeqnarraynumspace
  \end{IEEEeqnarray}
  where $P\sub{out}(R)$ is the outage probability, defined as
  \begin{IEEEeqnarray}{rCL}\label{eq:outage_probability}
    %\IEEEeqnarraymulticol{3}{l}{...}
    % a & = & b +c
    P\sub{out}(R)=\Pr\Biggl\{\sum_{l=1}^{L}\ln\det\Bigl(\matI+\frac{\rho}{\txant}\randmatH_l\herm{\randmatH_l}\Bigr)<R \Biggr\}.\IEEEeqnarraynumspace
  \end{IEEEeqnarray}
   The outage capacity does not depend on whether CSI is actually available at the receiver or not.
   Indeed, as the coherent interval~$\cohtime$ increases, the cost of learning the channel at the receiver vanishes~\cite{yang14-07c}.

   The diversity gain $d^*(r)$ as a function of the multiplexing gain~$r$ (as defined in~\cite{zheng03-05a}) is obtained by setting $R=r\ln \snr$ in~\eqref{eq:outage_probability}, and by computing the asymptotic ratio
  \begin{IEEEeqnarray}{rCL}
    %\IEEEeqnarraymulticol{3}{l}{...}
    % a & = & b +c
    d^*(r)=-\lim_{\snr\to\infty}\frac{P\sub{out}(r\ln \snr)}{\ln \snr}.
  \end{IEEEeqnarray}
  For the channel~\eqref{eq:channel_model}, the diversity $d^*(r)$ as a function of~$r$ is a piecewise linear curve joining the points $(k, L(\txant-k)(\rxant-k))$ where $k=1,\dots,\min\{\txant,\rxant\}$.
  The price to be paid for such an elegant characterization of the tradeoff between diversity and multiplexing is its asymptotic nature in both the blocklength and the SNR, which may limit its significance for the scenarios analyzed in this paper.
  
  For the case when $L$, and hence $n$, grows to infinity for a fixed~$T$, the maximum coding rate  $R^{*}(n,\epsilon)$ converges to the ergodic capacity $C_{\mathrm{erg}}(\snr)$, which, for the case when the constraint~\eqref{eq:code-constraint} is relaxed, is given by
  \begin{IEEEeqnarray}{rCL}\label{eq:ergodic_capacity}
    %\IEEEeqnarraymulticol{3}{l}{...}
    % a & = & b +c
    C_{\mathrm{erg}}(\snr)=({1}/{\cohtime})\sup I(\randmatX;\randmatY)
  \end{IEEEeqnarray}
  where $\randmatY=\sqrt{\rho/\txant}\randmatX\randmatH+\randmatW$ and the supremum is over all probability distributions of $\cohtime\times\txant$ matrices $\randmatX$ satisfying $\Ex{}{\tr{\randmatX\herm{\randmatX}}}\leq \cohtime\txant$.
  A closed-form expression for~\eqref{eq:ergodic_capacity} is not available. It is known that choosing  $\randmatX$ to be an isotropically distributed scaled unitary matrix  achieves the first two terms in the high-SNR expansion of $C_{\mathrm{erg}}(\snr)$ for the case when $\cohtime\geq \txant+\rxant$~\cite{zheng02-02a,yang13-02a}.
  Such an input distribution, which is often referred to as \emph{unitary space-time modulation (USTM)}, yields codewords satisfying~\eqref{eq:code-constraint}.
This result thus provides support for introducing the additional codeword constraint~\eqref{eq:code-constraint}.
In the remainder of the paper, we shall focus on the case $\cohtime\geq \txant+\rxant$ and denote the USTM input distribution by~$P_{\randmatX}^{\mathrm{U}}$.
% subsection outage_and_ergodic_capacity (end)

\section{Bounds on $R^*(n,\epsilon)$} % (fold)
\label{sec:bounds_on_r_n_epsilon_}
\subsection{Output distribution induced by USTM} % (fold)
\label{sec:output_distribution_induced_by_ustm}
A fundamental ingredient of the nonasymptotic bounds on $R^*(n,\epsilon)$ described in this section is the following closed-form expression for the probability density function (pdf) induced on the channel output $\randmatY=\sqrt{\rho/\txant}\randmatX\randmatH+\randmatW$ by a USTM-distributed input~$\randmatX\sim P_{\randmatX}^{\mathrm{U}}$.
\begin{prop}
\label{prop:output-pdf}
The pdf induced on $\randmatY$ in~\eqref{eq:channel_model} by an USTM input distribution $P_{\randmatX}^{\mathrm{U}}$ is 
\begin{IEEEeqnarray}{rCl}
\genericpdf_{\randmatY}(\matY)
 &=& \frac{\lefto(1 + \snr\cohtime/\txant\right)^{\txant(\cohtime - \txant-\rxant)} }{\pi^{\rxant\cohtime} (\snr\cohtime/\txant)^{\txant(\cohtime -\txant)} }\notag\\
 &&\cdot\frac{ \Gamma_{\txant}(\cohtime)}{  \Gamma_{\txant}(\txant)}  \funcpdf_{\snr,\txant,\cohtime}(\sigma_1^2,\ldots,\sigma_{\rxant}^2) .    \IEEEeqnarraynumspace
% \notag\\ &&\cdot\, . \IEEEeqnarraynumspace
 \label{eq:out-density-Y-USTM}
\end{IEEEeqnarray}
Here, $\sigma_1\geq \cdots \geq \sigma_{\rxant}$ denote the ordered singular values of $\matY$, and $\funcpdf_{\snr,\txant,\cohtime}: \realset^{\rxant}_{+} \to \realset_{+}$ is defined as
\begin{IEEEeqnarray}{rCl}
\funcpdf_{\snr,\txant,\cohtime}(b_1,\ldots,b_{\rxant}) &\define&
 \frac{ \det\matA}{\prod\nolimits^{\rxant}_{i<j}(b_i -b_j) } \notag \\
  &&\cdot\prod\limits_{i=1}^{\rxant}  \frac{ e^{-(1+\snr\cohtime/\txant)^{-1}b_i} }  {b_i^{\cohtime-\rxant}}
  \label{eq:def-func-in-pdf}
\end{IEEEeqnarray}
where $\matA  $ is an $\rxant\times\rxant$ real matrix whose $(j,k)$th entry is 
\begin{IEEEeqnarray}{rCl}
[\matA]_{jk}=
\begin{cases}
b_{k}^{\txant -j} \altgamma\lefto(\cohtime +j-2\txant, \frac{\snr\cohtime b_k}{\txant + \snr \cohtime}\right),& \, \text{ if } L\leq \txant\\
b_k^{\cohtime-j} \exp\lefto[-b_k\snr T/(\txant+\snr T)\right],& \, \text{otherwise}
\end{cases}
\IEEEeqnarraynumspace
\end{IEEEeqnarray}
with
\begin{equation}
\altgamma(n,x) \define \frac{1}{\Gamma(n)}\int\nolimits_{0}^{x}t^{n-1} e^{-t} dt
\end{equation}
denoting the regularized incomplete Gamma function.
%$\matF_{\snr,\txant,\cohtime}: \realset^{\txant} \to \realset^{\txant\times\txant}$ is defined in~\eqref{eq:def-matrix-function}.
\end{prop}

\begin{IEEEproof}
See~\cite{ostman14-08a}.
\end{IEEEproof}

\begin{rem}
A different expression for the output pdf induced by $\unifdist_{\randmatX}$ is reported in~\cite{hassibi02-06a}. 
The expression in Proposition~\ref{prop:output-pdf}  appears to be easier to compute and more stable numerically.
\end{rem}

% subsection output_distribution_induced_by_ustm (end)

\subsection{USTM Dependency Testing Lower Bound} % (fold)
\label{sec:dependency_testing_lower_bounds}

% subsection dependency_testing_lower_bounds (end)
We first present a lower bound on $R^*(n,\epsilon)$ that is based on the dependency testing (DT) bound developed in~\cite[Th.~17]{polyanskiy10-05a} and makes use of the USTM input distribution.
\begin{thm}
Let $\randmatZ_l $, $l=1,\ldots,\codeblocks$, be independent complex Gaussian $\cohtime\times \rxant$ matrices with i.i.d.\ $\jpg(0,1)$ entries.
Let $\Lambda_{l,1} \geq \ldots\geq \Lambda_{l,\rxant}$ denote the ordered eigenvalues of $\herm{\randmatZ} \matD \randmatZ$, where $\matD\in\realset^{\cohtime\times\cohtime}$ is given by
\begin{IEEEeqnarray}{rCl}
\matD =\mathrm{diag}\mathopen{}\Big\{\underbrace{1+\snr\cohtime/\txant,...,1+\snr\cohtime/\txant}_{\txant},\underbrace{1,...,1}_{\cohtime - \txant}\Big\}.
\IEEEeqnarraynumspace
\label{eq:def-diag-matrix}
\end{IEEEeqnarray}
Let
\begin{IEEEeqnarray}{rCl}
S_l &\define& c_{\txant,\cohtime} -   \,\tr{\herm{\randmatZ_l}\randmatZ_l} -
\ln \funcpdf_{\snr,\txant,\cohtime}(\Lambda_{l,1},\ldots,\Lambda_{l,\rxant}) \IEEEeqnarraynumspace
%
% \frac{\txant \ln e  }{\txant + \snr\cohtime}\! \sum\limits_{i=1}^{\txant}\! \Lambda_{l,i}   \notag\\
%&& +\, (\cohtime - \txant)\sum\limits_{i=1}^{\txant}\log \Lambda_{l,i}  + \sum\limits_{i<j}^{\txant}\log(\Lambda_{l,i}-\Lambda_{l,j}) \notag\\
%&&- \log \det \matF_{\snr,\txant,\cohtime}(\Lambda_1,\ldots,\Lambda_{\txant})
 \label{eq:def-S-l}
\end{IEEEeqnarray}
where
\begin{IEEEeqnarray}{rCl}
c_{\txant,\cohtime} \define \txant(\cohtime-\txant)\ln \frac{\snr \cohtime}{\txant + \snr\cohtime} - \ln \frac{\Gamma_{\txant}(\cohtime)}{\Gamma_{\txant}(\txant)}\IEEEeqnarraynumspace
\label{eq:def-c-t-T}
\end{IEEEeqnarray}
and $\funcpdf_{\snr,\txant,\cohtime}: \realset^{\rxant}_{+} \to \realset_{+}$ is defined in~\eqref{eq:def-func-in-pdf}.
%
%
%$\matF_{\snr,\txant,\cohtime}: \realset^{\txant} \to \realset^{\txant\times\txant}$ is a real-matrix-valued function whose $(j,k)$-th coordinate $f_{j,k} $ is defined as
%\begin{equation}
%f_{j,k} (x_1,\ldots,x_{\txant}) \define x_{k}^{\txant -j} \altgamma\lefto(\cohtime +j-2\txant, \frac{\snr\cohtime x_k}{\txant + \snr \cohtime}\right)
%\label{eq:def-matrix-function}
%\end{equation}
%
Furthermore, let
\begin{IEEEeqnarray}{rCl}
\error_{\mathrm{ub}} (\bl, \NumCode) \define
\Ex{}{\exp\lefto\{-\Biggl[\sum\limits_{l=1}^{\codeblocks}S_l -\ln\frac{\NumCode-1}{2}\Biggr]^{+}\right\}}. \IEEEeqnarraynumspace
\end{IEEEeqnarray}
We have
\begin{IEEEeqnarray}{rCl}
R^*(\bl,\error) \geq \max\lefto\{{(\ln \NumCode)}/{\bl} : \error_{\mathrm{ub}}(\bl, \NumCode) \leq \error\right\}.
\label{eq:lower-bound-DT-thm}
\end{IEEEeqnarray}
\end{thm}

\begin{IEEEproof}
The lower bound~\eqref{eq:lower-bound-DT-thm} follows by applying the DT bound~\cite[Th.~17]{polyanskiy10-05a} with an input distribution $\unifdist_{\randmatX^{\codeblocks}}$ chosen so that the subcodewords $\{\randmatX_{l}\}_{l=1}^\codeblocks$, are i.i.d. $P_{\randmatX}^{\mathrm{U}}$-distributed.
Let $\genericpdf_{\randmatY^\codeblocks}$  denote the pdf of  $\randmatY^\codeblocks$ induced by~$\unifdist_{\randmatX^{\codeblocks}}$,
and $\genericpdf_{\randmatY^L \given \randmatX^L}$ denote the conditional pdf of~$\randmatY^\codeblocks$ given $\randmatX^{\codeblocks}$.
To use the DT bound~\cite[Th.~17]{polyanskiy10-05a}, we need to compute the information density
\begin{IEEEeqnarray}{rCl}
\infoden_{\codeblocks}\lefto(\matX^{\codeblocks} ; \matY^{\codeblocks}\right) \define \ln \frac{\genericpdf_{\randmatY^L \given \randmatX^L} (\matY^\codeblocks \given \matX^\codeblocks)}{ \genericpdf_{\randmatY^\codeblocks}(\matY^\codeblocks)}.
\label{eq:def-info-density}
\end{IEEEeqnarray}
Since the channel~\eqref{eq:channel_model} is block-memoryless, $\genericpdf_{\randmatY^{\codeblocks} \given \randmatX^{\codeblocks}}(\matY^{\codeblocks} \given \matX^{\codeblocks})$ factorizes as $\prod_{l=1}^{\codeblocks} \genericpdf_{\randmatY \given \randmatX}(\matY_l \given \matX_l)$ with
\begin{IEEEeqnarray}{rCl}
\genericpdf_{\randmatY\given \randmatX }(\matY \given \matX) = \frac{e^{-\tr{\herm{\matY} \bigl(\snr\txant^{-1} \matX\herm{\matX} + \matI_{\cohtime} \bigr)^{-1} \matY} } }{\pi^{\rxant\cohtime} \det(\snr\txant^{-1} \matX\herm{\matX} +\matI_{\cohtime})^{\rxant}}.
\end{IEEEeqnarray}
Moreover, since the $\{\randmatX_{l}\}_{l=1}^{\codeblocks}$ are i.i.d., $\genericpdf_{\randmatY^L }$ also factorizes as
\begin{IEEEeqnarray}{rCl}
\genericpdf_{\randmatY^L} (\matY^L)= \prod_{l=1}^{\codeblocks} \genericpdf_{\randmatY}(\matY_l)
\end{IEEEeqnarray}
where $ \genericpdf_{\randmatY}$ is given in~\eqref{eq:out-density-Y-USTM}.
It thus follows that
\begin{IEEEeqnarray}{rCl}
\infoden_{\codeblocks}(\matX^{\codeblocks},\matY^{\codeblocks}) = \sum\limits_{l=1}^{\codeblocks} \infoden (\matX_l ; \matY_l)
\end{IEEEeqnarray}
where
\begin{IEEEeqnarray}{rCl}
\infoden (\matX ; \matY) = \ln \frac{\genericpdf_{\randmatY\given \randmatX} (\matY \given \matX)}{ \genericpdf_{\randmatY}(\matY)}.
\end{IEEEeqnarray}
%
%A closed-form expression for $\genericpdf_{\randmatY}$ is provided in the following proposition.

%To prove~\eqref{eq:lower-bound-DT-thm} It suffices to show that under $P_{\randmatY}^{\mathrm{U}}$

%\todo{Maybe state this proposition in Section~\ref{sec:system_model}.}

%We next study the probability distribution of the random variable $\infoden_{\codeblocks}\lefto(\matX^{\codeblocks} ; \randmatY^{\codeblocks}\right)$ where $\randmatY^{\codeblocks}$ is distributed according to $P_{\randmatY^L\given \randmatX^L = \matX^\codeblocks}$.
%
%
%
Since $\unifdist_{\randmatY}$ is isotropic, the probability distribution of $\infoden\lefto(\matX_l ; \randmatY_l\right)$ takes the same value for all subcodeword matrices $\matX_l$ satisfying~\eqref{eq:code-constraint}.
Without loss of generality, we choose $\matX_l  = \bar{\matX}$, $l=1,\ldots,\codeblocks$, where
\begin{IEEEeqnarray}{rCl}
\bar{\matX} \define \left[
             \begin{array}{c}
               \sqrt{T} \matI_{\txant} \\
               \mathbf{0} \\
             \end{array}
           \right].
           \label{eq:def-x-bar}
\end{IEEEeqnarray}
Through algebraic manipulations, one can show that, under $P_{\randmatY^\codeblocks\given \randmatX^\codeblocks = \bar{\matX}^\codeblocks}$, the random variable $\infoden\lefto(\bar{\matX}_l ; \randmatY_l\right)$ has the same distribution as $S_{l}$. 
Hence, $\infoden_{\codeblocks}\lefto(\bar{\matX}^{\codeblocks} ; \randmatY^{\codeblocks}\right)$ has the same distribution as $\sum\nolimits_{l=1}^{\codeblocks} S_{l}$.
%
%The proof is concluded by noting that under $P_{\randmatX^L}^{\mathrm{U}} P_{\randmatY^L\given \randmatX^L}$ is distributed as $S_{}$
%Applying the DT bound~\cite[Th.~17]{polyanskiy10-05a}, we conclude that there exists a code with $\NumCode$ codewords and average probability of error not exceeding $\error_{\mathrm{ub}}(\bl,\NumCode)$. 
This proves~\eqref{eq:lower-bound-DT-thm}.
\end{IEEEproof}

% subsubsection m (end)

% \subsection{Transmit diversity/spatial multiplexing lower bounds} % (fold)
% \label{sec:transmit_diversity_lower_bound}
% %
% \todo{Maybe omit this section all together and describe these two architectures in the section about simulation results }
% %
% We next consider two specific transceiver architectures, one exploiting transmit diversity, the other spatial multiplexing and develop corresponding lower bounds on $R^*(\bl,\error)$ for these two architectures based on the DT bound.
% %
% \todo{Transmit diversity: STOC-Alamouti; channel estimation at the receiver (if we want to get there)}
% %
% \todo{Spatial multiplexing: ZF receiver that decodes each single stream. Equal rate allocation across streams. Outage if any of the streams is not decoded correctly.}
% \begin{IEEEeqnarray}{rCL}
%   %\IEEEeqnarraymulticol{3}{l}{...}
%   % a & = & b +c
%   Y_l= 
% \end{IEEEeqnarray}
% %
% and thus exploits the available $\txant\rxant$ diversity branches
% % subsection transmit_diversity_lower_bound (end)
% \todo{Insert theorem: proof: same as ITW;}
% 
% \todo{Transmit diversity and spatial multiplexing lower bound: insert rate sacrificing diversity scheme that transforms the MIMO channel into a SISO channel to exploit the available $\txant\rxant$ diversity branches. Resulting input-output relation}

\subsection{Meta-converse Upper Bound} % (fold)
\label{sec:meta_converse_upper_bound}
%\todo{Follows essentially from~\cite{yang14-03a}}
% subsection meta_converse_upper_bound (end)

We next establish an upper bound on~$R^*(\bl,\error)$.
\begin{thm}
Let the random variables $\{\bar{\randmatY}_l \in \complexset^{\cohtime \times \rxant}\}$, $l=1,\ldots,\codeblocks$, be i.i.d.\ $\unifdist_{\randmatY}$-distributed; let $\Delta_{l,1},\ldots,\Delta_{l,\rxant}$ denote the nonzero singular values of $\bar{\randmatY}_l$, and let $\randmatDelta_l \define \diag{ \Delta_{l,1},\ldots,\Delta_{l, \rxant}}$.
Further let $\{\randmatU_l \in \complexset^{\cohtime \times \txant}\}$, $l=1,\ldots,\codeblocks$, be i.i.d.\ isotropically distributed unitary matrices;
let~$\matD$,~$S_{l}$,~$ c_{\txant,\cohtime}$, and~$\funcpdf_{\snr,\txant,\cohtime}$ be defined as in~\eqref{eq:def-diag-matrix},~\eqref{eq:def-S-l},~\eqref{eq:def-c-t-T}, and~\eqref{eq:def-func-in-pdf}, respectively.
Finally, let
\begin{IEEEeqnarray}{rCl}
K_{l} &\define& c_{\txant,\cohtime} - \tr{\randmatDelta_l^2 \herm{\randmatU}_l \matD^{-1} \randmatU_l} \notag\\
&&  - \, \ln \funcpdf_{\snr,\txant,\cohtime}(\Delta_{l,1},\ldots, \Delta_{l,\rxant}).
%\frac{\txant \log e  }{\txant + \snr\cohtime} \sum\limits_{i=1}^{\txant} \Delta_{l,i}\notag\\
%&& +\, (\cohtime - \txant)\sum\limits_{i=1}^{\txant}\log \Delta_{l,i}  + \sum\limits_{i<j}^{\txant}\log(\Delta_{l,i}-\Delta_{l,j}) \notag\\
%&& -\, \log \det\matF_{\snr,\txant,\cohtime}(\Delta_1,\ldots, \Delta_{\txant}) .
\end{IEEEeqnarray}
Then, for every $\bl$ and every $0<\error<1$, the maximal channel coding rate $R^*(\bl,\error)$ for the block-memoryless Rayleigh-fading channel~\eqref{eq:channel_model} is upper-bounded by
\begin{IEEEeqnarray}{rCl}\label{eq:MC_UB}
R^*(\bl,\error) \leq \frac{1}{\bl}\ln \frac{1}{\prob\mathopen{}\big[\sum\nolimits_{l=1}^{\codeblocks} K_l \geq \bl\gamma_\bl\big]}
\end{IEEEeqnarray}
where $\gamma_\bl$ is the solution of
\begin{IEEEeqnarray}{rCl}
\prob\mathopen{}\biggl[\sum_{l=1}^{\codeblocks} S_l \leq \bl\gamma_\bl\biggr] =\error.
\end{IEEEeqnarray}
 \end{thm}
\begin{rem}
The conditional probability $\prob\mathopen{}\big[\sum\nolimits_{l=1}^{\codeblocks} K_l \geq \bl\gamma_\bl | \{\randmatDelta_{l}\} \big]$ can be computed in closed-form by evaluating the conditional characteristic function of $K_{l}$ given $\randmatDelta_{l}$ using the Itzykson-Zuber integral formula~\cite[Eq.~(3.4)]{itzykson80-a}.
\end{rem}

\begin{IEEEproof}
To upper-bound~$R^*(\bl,\error)$, we use the meta-converse (MC) bound~\cite[Th.~28]{polyanskiy10-05a} with the following auxiliary distribution
\begin{IEEEeqnarray}{rCl}
Q_{\randmatY^{\codeblocks}} = \prod\limits_{l=1}^{\codeblocks} \unifdist_{\randmatY}.
\end{IEEEeqnarray}
 Since $\unifdist_{\randmatY}$ is isotropically distributed, $\beta_{1-\error}(P_{\randmatY^{\codeblocks}\given \randmatX^{\codeblocks} =  \matX^{\codeblocks}}, Q_{\randmatY^{\codeblocks}})$ (as defined in \cite[p.~2316]{polyanskiy10-05a}) takes the same value for all $\matX^{\codeblocks}$ satisfying~\eqref{eq:code-constraint}.
Thus, the MC bound yields \cite[Th.~28]{polyanskiy10-05a}
\begin{IEEEeqnarray}{rCl}
R^*(\bl,\error) \leq \frac{1}{\bl}\ln \frac{1}{\beta_{1-\error}(P_{\randmatY^{\codeblocks}\given \randmatX^{\codeblocks} =  \bar{\matX}^{\codeblocks}}, Q_{\randmatY^{\codeblocks}} )}
\end{IEEEeqnarray}
where $\bar{\matX}^{\codeblocks}=[\bar{\matX},\dots,\bar{\matX}]$, with $\bar{\matX}$ given in~\eqref{eq:def-x-bar}. By the Neyman-Pearson lemma,
\begin{IEEEeqnarray}{rCl}
\beta_{1-\error}(P_{\randmatY^{\codeblocks}\given \randmatX^{\codeblocks} =  \bar{\matX}^{\codeblocks}}, Q_{\randmatY^{\codeblocks}} ) = Q_{\randmatY^{\codeblocks}}[\infoden_L(\bar{\matX}^{\codeblocks}; \randmatY^{\codeblocks}) \geq \bl \gamma_\bl] \IEEEeqnarraynumspace
\end{IEEEeqnarray}
where $\infoden_L(\cdot,\cdot)$ is defined in~\eqref{eq:def-info-density} and $\gamma_\bl$ is the solution of
\begin{IEEEeqnarray}{rCl}
P_{\randmatY^{\codeblocks}\given \randmatX^{\codeblocks} =  \bar{\matX}^{\codeblocks}}
\mathopen{}\big[ \infoden_L(\bar{\matX}^{\codeblocks}; \randmatY^{\codeblocks}) \leq \bl\gamma_\bl\big] =\error.
\end{IEEEeqnarray}
We conclude the proof by noting that, under $Q_{\randmatY^{\codeblocks}}$, the random variable $\infoden_L(\bar{\matX}^{\codeblocks}; \randmatY^{\codeblocks}) $ has the same distribution as $\sum\nolimits_{l=1}^{\codeblocks} K_l$, and under $P_{\randmatY^{\codeblocks}\given \randmatX^{\codeblocks} =  \bar{\matX}^{\codeblocks}}$, it has the same distribution as $\sum\nolimits_{l=1}^{\codeblocks} S_l $.
\end{IEEEproof}

% section bounds_on_r_n_epsilon_ (end)

% section system_model (end)

\section{Numerical Results} % (fold)
\begin{figure}[t]
  \centering
    \includegraphics[width=\figwidth]{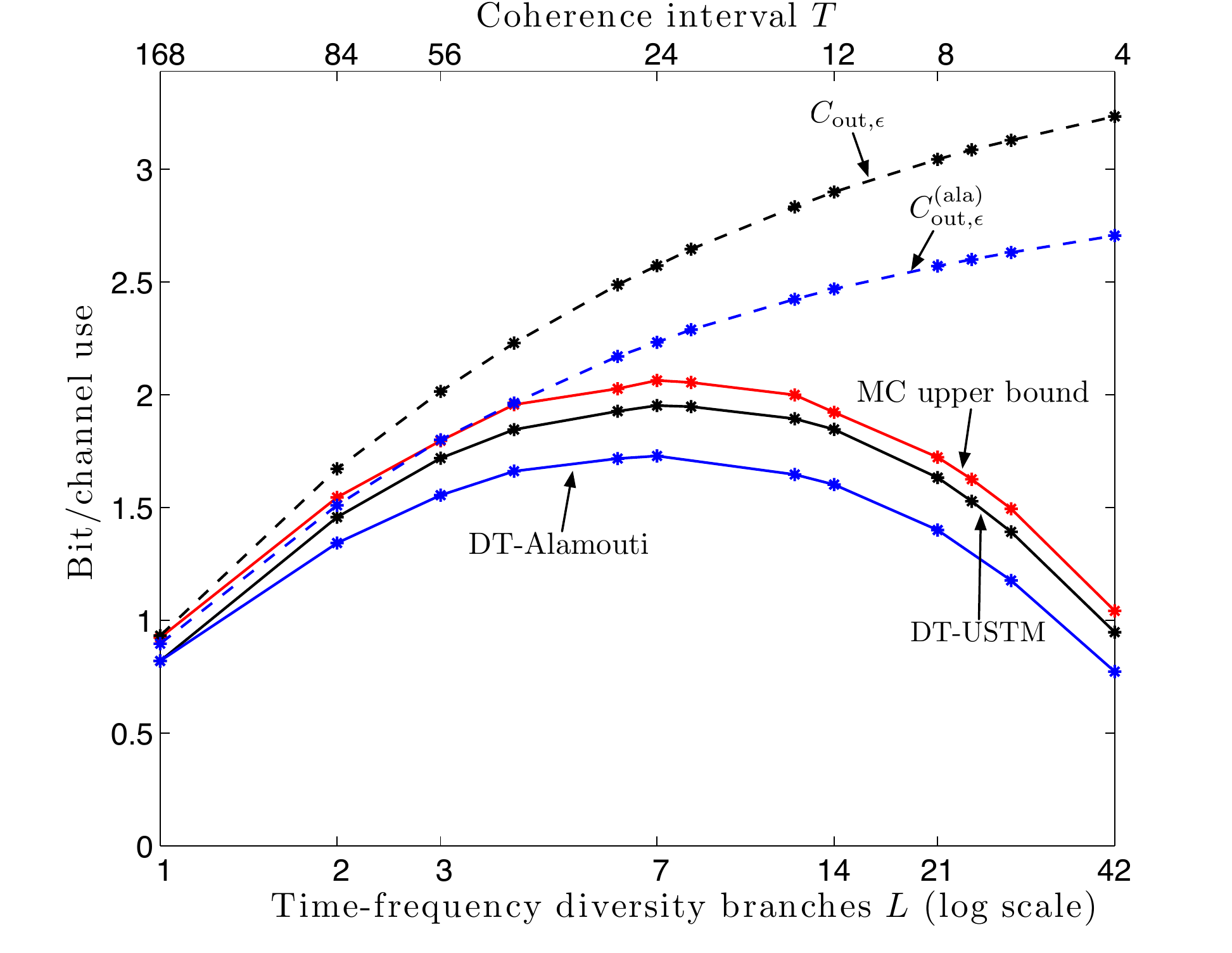}
  \caption{$\txant=\rxant=2$, $n=168$, $\epsilon=10^{-3}$.}
  \label{fig:figs_figs_final_2x2_1e-3}
\end{figure}

\label{sec:numerical_results}
We consider the same setup as in~\cite{lozano10-09a}, which is based on the 3GPP LTE standard~\cite{sesia11}. 
Specifically, the packet length is set to $n=168$ symbols (an LTE resource block).
Throughout, we set $\snr=6$ dB.
\paragraph*{Control signaling} % (fold)
\label{par:control_information_in_lte_}
We first consider the case where the packet error rate is $\epsilon=10^{-3}$, which may be appropriate for the exchange of short packets carrying control signaling.
In Fig.~\ref{fig:figs_figs_final_2x2_1e-3}, we plot the DT lower bound~\eqref{eq:lower-bound-DT-thm} and the MC upper bound~\eqref{eq:MC_UB} for the case $\txant=\rxant=2$ as a function of the coherence interval $T$.
% \begin{figure}[t]
%   \centering
%     \includegraphics[width=.8\figwidth]{figs/2x2_SNR=6dB_n=168-eps-converted-to.pdf}
%   \caption{\todo{write caption and replace with final version when ready; it is probably a good idea to have two x-axis, one for $L$, the other for $T$.  The formula for outage capacity depends on $L$ only: there is no $T$ or $n$. Also the Alamouti outage capacity needs to be fixed and Alamouti-DT added. Good to have same scale for all plots. It facilitates comparisons.}}
%   \label{fig:figs_2x2_SNR=6dB_n=168-eps-converted-to}
% \end{figure}
% 
\begin{figure}[t]
  \centering
    \includegraphics[width=\figwidth]{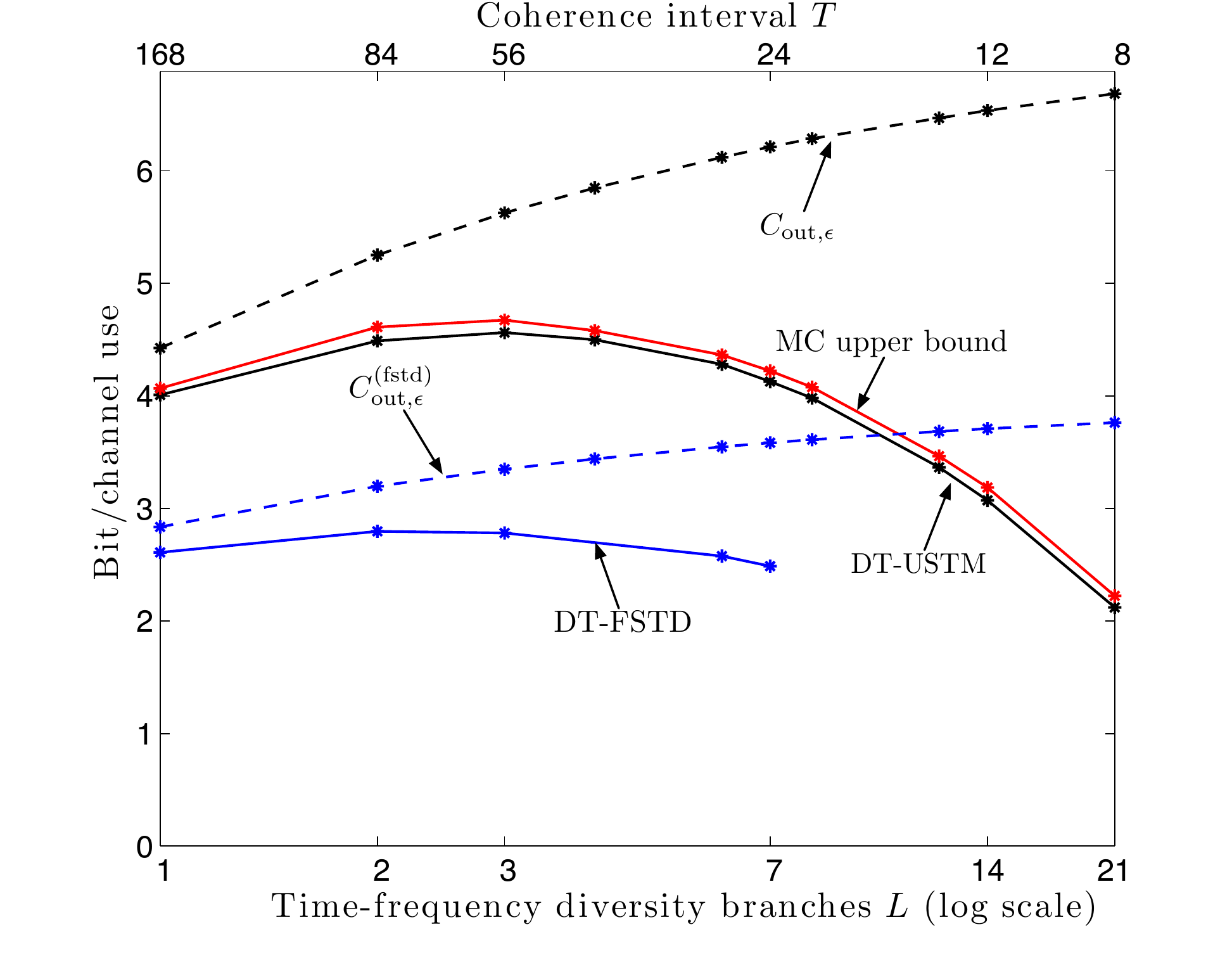}
  \caption{$\txant=\rxant=4$, $n=168$, $\epsilon=10^{-3}$.}
  \label{fig:figs_figs_final_4x4_1e-3}
\end{figure}
These bounds describe $R^*(n,\epsilon)$ accurately and demonstrate that $R^*(n,\epsilon)$ is not monotonic in the coherence interval $T$, but that there exists an optimal value $T^*$ (or equivalently, an optimal number $L^*=n/T^*$ of time-frequency diversity branches) that maximizes it. 
For $T<T^*$, the cost of estimating the channel overcomes the gain due to time-frequency diversity. 
For $T>T^*$, the bottleneck is the limited time-frequency diversity offered by the channel.  
A similar observation was reported in~\cite{yang12-09a} for the single-antenna case.
%The fact that USTM-based bounds describe the behavior of $R^*(n,\epsilon)$ accurately 

In the figure, we also plot the outage capacity $C_{\mathrm{out},\epsilon}$ in~\eqref{eq:outage_capacity} as a function of the number of time-frequency diversity branches $L=n/T$ (with $n=168$). 
As shown in the figure, the outage capacity provides a good approximation for $R^*(n,\epsilon)$ only when $T\approx n$, i.e., when the fading channel is essentially constant over the duration of the packet (quasi-static scenario).
Furthermore, $C_{\mathrm{out},\epsilon}$ fails to capture the loss in throughput due to channel estimation overhead, which is relevant for small $T$. 

The two additional curves in Fig.~\ref{fig:figs_figs_final_2x2_1e-3} correspond to
the case of Alamouti transmission~\cite{alamouti98-10a}, a scheme that provides diversity gain $4$ but no multiplexing gain~\cite[Sec.~9.1.5]{tse05a}. 
Both a DT bound (not detailed here for space limitations) and the outage-versus-achievable rate $C^{(\mathrm{ala})}_{\mathrm{out},\epsilon}$ of this scheme are depicted.
Fig.~\ref{fig:figs_figs_final_2x2_1e-3} demonstrates that, if outage capacity is used as performance metric, then a diversity-exploiting scheme such as the Alamouti code is nearly optimal when the channel provides limited diversity in time and frequency ($L\approx 1$). However, if the channel provides significant time-frequency diversity, then one should use the antennas in multiplexing mode. For example, for the case $L=14$, the gap between the Alamouti scheme and the outage-optimal scheme is about $0.5$ bit/channel use. 
Furthermore, this gap increases as $L$ grows.
This observation is one of the key contributions of~\cite{lozano10-09a}.

The picture changes when the limited packet size is accounted for. When $L$ is large, and hence $T$ is small, the cost of learning the channel is significant.
This means that large multiplexing gains are not feasible and the gap to optimality of the Alamouti scheme decreases. 
For example, the throughput reduction due to the use of the Alamouti scheme (evaluated comparing the DT-Alamouti and the DT-USTM lower bounds) is about $0$ bit/channel use for $L=1$; $0.25$ bit/channel use  for $L=12$; and $0.18$ bit/channel use  for $L=42$.

In Fig.~\ref{fig:figs_figs_final_4x4_1e-3}, we present a similar comparison for the case of a $4\times 4$  system.
As no generalization of the Alamouti scheme exists beyond the $2\times 2$ configuration~\cite{tarokh99-07a}, we consider instead the combination of Alamouti and  Frequency Switched Transmit Diversity (FSTD) used in LTE~\cite[Sec.~11.2.2.1]{sesia11}. This scheme provides diversity gain $8$ and no multiplexing gain.
As shown in the figure, the gap between the MC upper bound and the DT-USTM lower bound is small, allowing for a precise characterization of $R^*(n,\epsilon)$. In contrast, the gap between the DT-USTM and the DT-FSTD lower bound is large, which suggests that using all $8$ antennas to provide diversity gain is suboptimal also when the time-frequency diversity is limited (i.e., $L$ is small).

\begin{figure}[t]
  \centering
    \includegraphics[width=\figwidth]{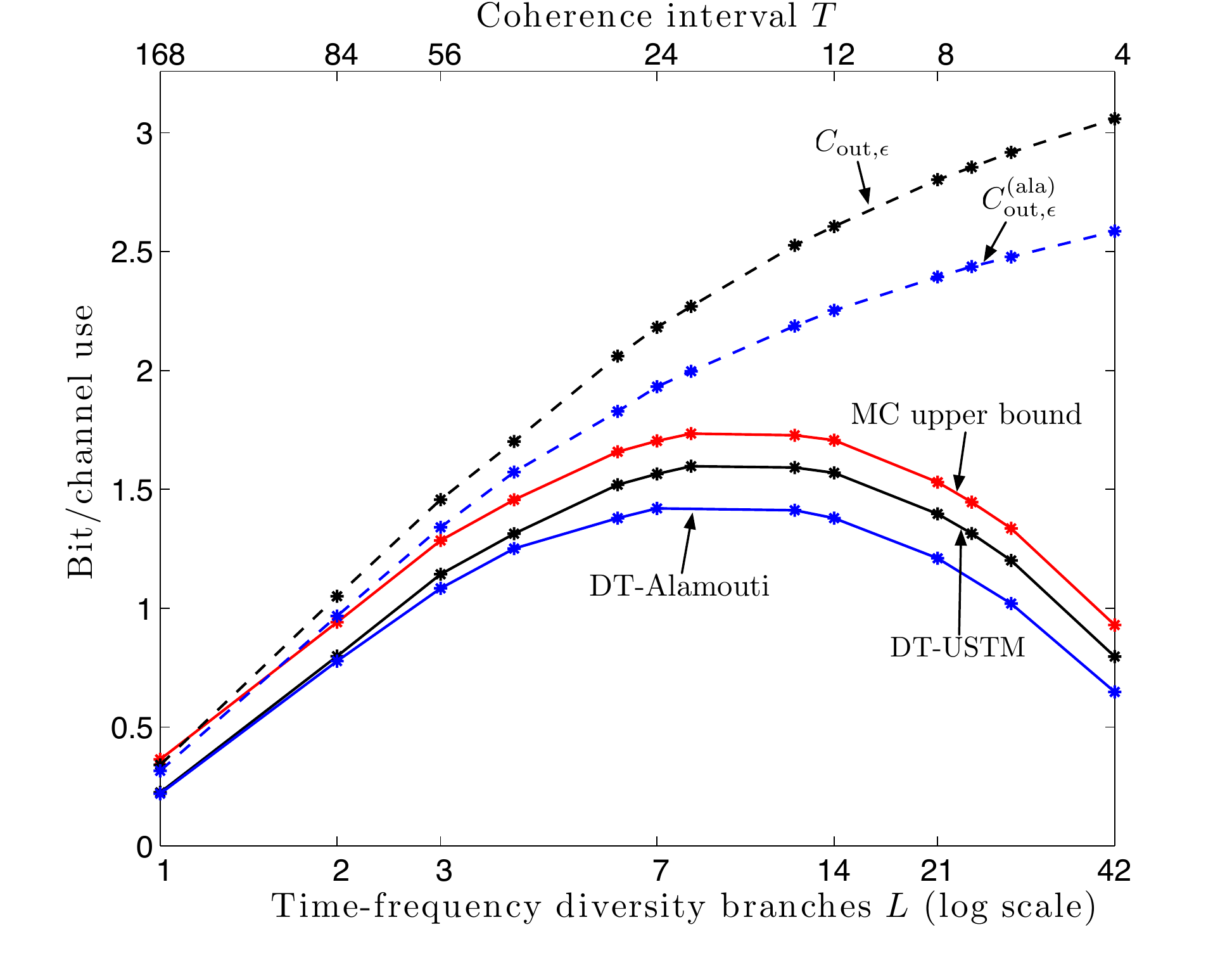}
  \caption{$\txant=\rxant=2$, $n=168$, $\epsilon=10^{-5}$.}
  \label{fig:figs_figs_final_2x2_1e-5}
\end{figure}

\begin{figure}[t]
  \centering
    \includegraphics[width=\figwidth]{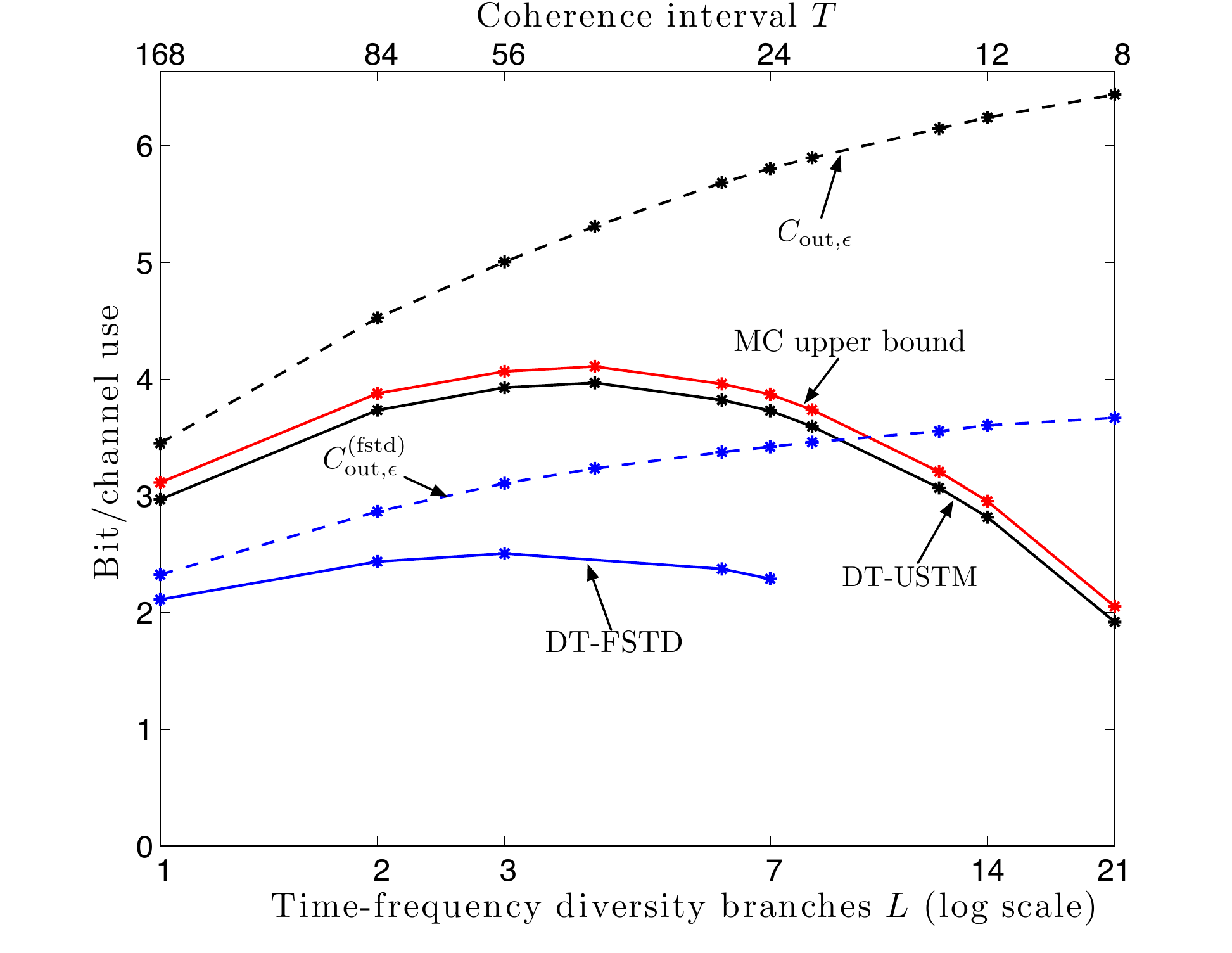}
  \caption{$\txant=\rxant=4$, $n=168$, $\epsilon=10^{-5}$.}
  \label{fig:figs_figs_final_4x4_1e-5}
\end{figure}

\paragraph*{Ultra reliable communication} % (fold)
\label{par:ultra_highly_reliable_communication}
In Fig.~\ref{fig:figs_figs_final_2x2_1e-5} and Fig.~\ref{fig:figs_figs_final_4x4_1e-5}, we consider the case $\epsilon=10^{-5}$, which may be relevant for the transmission of critical information, e.g., in traffic-safety applications~\cite{metis-project-deliverable-d1.113-04a}.
%
%\vspace*{-1mm}
Compared to the case $\epsilon=10^{-3}$, the gap between the optimal schemes and the diversity-based schemes (Alamouti for the $2\times 2$ configuration, and FSTD for the $4\times 4$ case) gets smaller. 
This comes as no surprise, since the higher reliability requirement makes the exploitation of transmit diversity advantageous.

% \section{Conclusions} % (fold)
% \label{sec:conclusions}
% The bounds on the maximum coding rate $R^{*}(n,\epsilon)$ reported in this paper allow one to assess for  which packet size, number of antennas, and degree of channel selectivity, diversity-exploiting schemes are close to optimal, and when instead the available spatial degrees of freedom should be used to provide spatial multiplexing.
% This finite-blocklength view on the diversity-multiplexing tradeoff provides insights on the design of delay-sensitive ultra-reliable communication
% links.
% section conclusions (end)

%%%%%%%%%%%%%%%%%%%%%%%%%%%%
\bibliographystyle{IEEEtran}
\bibliography{IEEEabrv,publishers,confs-jrnls,giubib}
\end{document}